# On the accuracy of the Debye shielding model

**M. A. Martínez-Fuentes[*] and J. J. E. Herrera-Velázquez**
Instituto de Ciencias Nucleares, Universidad Nacional Autónoma de México
Apdo. Postal 70-543, Ciudad Universitaria, 04511 México, D.F., México
* Present address: Laboratorio Nacional de Fusión, Ciemat, Madrid, Spain.

e-mail: herrera@nucleares.unam.mx

**Abstract.** The expression for the Debye shielding in plasma physics is usually derived under the assumptions that the plasma particles are weakly coupled, so that their total kinetic energy is much greater than their electrostatic interaction energy, and that the velocity distributions of the plasma species are Maxwellian. The first assumption also establishes that the number of particles within a sphere with a Debye radius, known as the plasma parameter $N_D$, should be significantly greater than 1, and determines the difference between weakly and strongly coupled plasmas. Under such assumptions, Poisson's equation can be linearized, and a simple analytic expression is obtained for the electrostatic potential. However, textbooks rarely discuss the accuracy of this approximation. In this work we compare the linearized solution with a more precise numerical (or "exact") solution, and show that the linearization, which underestimates the "exact" solution, is reasonably good even for $N_D \sim 40$. We give quantitative criteria to set the limit of the approximation when the number of particles is very small, or the distance to the test charge too short.

**1. Introduction**
One of the most elementary concepts in classical plasma physics is that of Debye shielding, which establishes the characteristic distance $\lambda_D$ in which the electrostatic field from a charged test particle can be shielded by particles of the opposite sign. If a test charge is placed in a homogenous plasma, in which the unperturbed electron and ion densities are the same (i.e.: satisfy the quasi-neutrality condition), it repels the charges of its own sign, while it "dresses" itself with charges of the opposite sign, resulting in an electrostatic shielding. The derivation of $\lambda_D$ has been well presented with similar arguments in many textbooks, such as Refs. [1-7]. In doing so, two assumptions are generally made: (i) The total kinetic energy of the particles is much greater than the electrostatic interaction energy between them, which means that the plasma is weakly coupled, and allows the linearization of the problem, (ii) Each particle species is in thermodynamic equilibrium, so their velocity distributions are Maxwellian. The latter assumption requires the density to be large enough to allow collisions to an equilibrium state, and does not apply for collision-less, non-equilibrium plasmas, such as those often found in space plasmas.

Modern textbooks discuss the case of strongly coupled plasmas, for which the first constraint does not apply [3, 5, 8], although they usually ignore the fact that many collision-less plasmas, particularly in space physics, even in steady state, are non-equilibrium systems, and therefore not Maxwellian. The case of shielding in quantum plasmas, in which Fermi statistics should be considered, has been presented in Ref. 8, but it will not concern us in this paper.

The question on the relaxation of assumption (ii) has been addressed by Bryant [9], who studied the shielding produced when the species are described by the so called kappa-distributions for non-equilibrium plasmas. This is a more general family of distributions that include the Maxwell-Boltzmann one as a special case, and is motivated by observations in space plasmas. While Bryant's conclusion is that the shielding $\lambda_D$ is shorter than for the Maxwellian case, given the same number density and mean energy, the weakly coupled assumption is kept. More recently, Livadiotis and Comas have reviewed the subject, providing a better theoretical ground for the kappa-distributions, under the point of view of non-extensive entropy [10]. At a didactical level Meyer-Vernet [11] studied the case in which a Boltzmann equilibrium is not assumed, from a different point of view.

On the other hand, the weakly coupled plasma assumption, also poses the requirement that the number of particles $N_D$ within a sphere of radius $\lambda_D$ be much greater than 1, so the case of classical strongly coupled plasmas, in which $N_D \sim 1$ (in the opposite limit of that of quantum plasmas, for which $N_D \gg 1$) is left open. It is also important to note that the linearized solution for the electrostatic potential, which results from the weakly coupled plasma assumption, is also limited in space, and breaks down for $r < \lambda_D$. The relaxation of the weakly coupled plasma assumption has been addressed by Mak [12], who compared the usual linearized solution for the electrostatic potential with the exact solution of Poisson's equation for several cases in which $N_D \gg 1$. While the conclusion was that the approximate solution is good beyond expectation for the cases studied, a few questions went unanswered. Particularly, when does the approximate solution breaks down as $N_D$ and the distance $r$ are reduced?

The purpose of this paper is to provide a better insight into the subject, using the same normalised approach as in Ref. [12], for comparison. In section 2 we go through the usual discussion of the problem, reviewing what has been presented in textbooks, for the sake of completeness and in order to establish the notation. In section 3 we compare the approximate analytical solution to the linearized problem with the exact numerical one, and in section 4 we summarise the main conclusions. The paper is written with sufficient detail, for easy reading by an advanced undergraduate or graduate physics student.

**2. Linearized Debye shielding**

*2.1 The Poisson equation*
Let us consider a homogeneous plasma, in which the unperturbed large-scale average electron and ion densities $n_{eo}$, and $n_{io}$, are equal: $n_{eo} = n_{io} = n$. If a point test charge is introduced at the origin, it will create an electrostatic potential $\Phi(r)$, which only depends on the radial coordinate $r$, and which must satisfy Poisson's equation

$$\frac{1}{r^2}\frac{d}{dr}\left(r^2 \frac{d\Phi(r)}{dr}\right) = -\frac{\rho_c(r)}{\varepsilon_o} \quad , \tag{2.1a}$$

where $\rho_c(r)$ is the charge density distribution. We make the usual assumptions (i) and (ii) mentioned in the introduction, and following Ref. [12], we make an additional assumption: (iii) The positive ions are protons with infinite inertia, so they form a uniform background of density $n$. Since the electron density would then be given by $n_e(r) = n \exp(e\Phi(r)/kT)$, where $-e$ is the charge of the electron, $k$ is Boltzmann's constant, and $T$ is the electron temperature, then $\rho_c(r) = ne[1 - \exp(e\Phi(r)/kT)]$, and equation (2.1a) can be rewritten as

$$\frac{d^2}{dr^2}[r\Phi(r)] = -\frac{ner}{\varepsilon_o}[1 - \exp(e\Phi(r)/kT)] \quad . \tag{2.1b}$$

The numerical solution to this equation is what we shall call the "exact solution".

## 2.2 The limit between weakly and strongly coupled plasmas

In this subsection we follow the discussion of Refs. [6, 7], with some adaptations. The boundary between weakly and strongly coupled plasmas can be established when the electrostatic energy between the plasma particles is equal to the kinetic energy between them, so weakly coupled plasmas are those for which the former is smaller than the latter, and the strongly coupled plasmas are represented by the opposite case. If we consider an electron with velocity $v$ and charge $-e$ in the presence of an ion with charge $Ze$ at rest, the distance of maximum approach $r_c$ will be defined by

$$\frac{1}{2}mv^2 - \frac{Ze^2}{4\pi\varepsilon_o r_c} = 0 \qquad . \qquad (2.2)$$

Taking $v$ as the thermal velocity $v_{th}$, in the case of three dimensions for an isotropic plasma, we can write $\frac{1}{2}mv_{th}^2 = \frac{3}{2}kT$. Since $\Phi(r) = \frac{Ze}{4\pi\varepsilon_o r_c}$,

$$\frac{e\Phi(r)}{kT} = \frac{3}{2} \qquad , \qquad (2.3)$$

which can be taken as the limiting case in Poisson's equation, such that weakly coupled plasmas happen for the right hand side is smaller than 3/2, and the opposite case for strongly coupled plasmas. For a weakly coupled plasma,

$$e\Phi(r)/kT \ll 1, \qquad (2.5)$$

in which we may take 1 instead of 3/2. Let us now define the Debye length $\lambda_D \equiv \sqrt{\varepsilon_o kT/ne^2}$ [1-7], whose significance will become clear in the linearized case. By taking the average distance between particles $r_d = n^{-1/3}$, (2.2) translates into

$$\frac{r_d}{r_c} = \frac{12\pi}{Z}\frac{\lambda_D^2}{r_d^2} \qquad , \qquad (2.6)$$

from which it becomes clear that weakly coupled plasmas are those for which both $r_d$ and $\lambda_D$ are greater than the maximum approach distance $r_c$. Using this result, it is easy to find that, if we define $\Lambda = \lambda_D/r_c$, as the ratio between the Debye length (the distance in linearized theory at which the electrostatic potential decays to $e^{-1}$ in the linear theory studied in the next subsection,) and the distance of maximum approach, we find that in the limit between weakly and strongly coupled plasmas, $\Lambda = (3/Z)N_D$, where $3N_D \equiv 4\pi n\lambda_D^3$ is the number of particles within a sphere defined by the Debye length (This definition for $N_D$ is for the sake of comparison with Ref. [12].) The familiar statement that in weakly coupled plasmas $N_D > 1$, while the opposite is true for strongly coupled plasmas, is recovered, as well as $(Z/3)\Lambda > 1$, for which the limit increases linearly with $Z$. All follow from assumption (2.5).

## 2.3 The weakly coupled limit

The linearization for the weakly coupled plasmas arises after considering $\Phi(r)/kT \ll 1$, which states the potential energy between the particles is much smaller than their kinetic energy, and is equivalent to $N_D \sim n\lambda_D^3 \gg 1$. This allows an expansion of the exponential in (2.1b). Keeping the first two terms this yields for the approximate potential $\Phi_a(r)$

$$\frac{d^2}{dr^2}[r\Phi_a(r)] = \frac{r\Phi_a(r)}{\lambda_D^2} \quad , \tag{2.7}$$

whose solution is the well known Yukawa potential:

$$\Phi_a(r) = A\frac{\exp(-r/\lambda_D)}{r} \quad . \tag{2.8}$$

By taking the integration constant $A = Ze/(4\pi\varepsilon_o)$, this allows the usual interpretation of the potential produced by a point charge of an ion $Ze$, with atomic number $Z$, at the origin ($\rho_c(r) = Ze\delta(r)$), where $\delta(r)$ is the Dirac delta function), shielded by the electrons with a characteristic decay length $\lambda_D$. Once the electrostatic potential is known, the charge distribution for the linearized solution can be obtained from equation (2.1a)

$$\rho_c(r) = \frac{Ze}{4\pi}\left[4\pi\delta(r) - \frac{\exp(-r/\lambda_D)}{r\lambda_D^2}\right] \quad . \tag{2.9}$$

The subtlety in obtaining (2.10) is discussed in Appendix. The effect of the shielding can be appreciated by computing the charge within a sphere of radius $r$:

$$Q(r) = Ze\left[1 - \int_0^r r'^2 \rho_c(r')dr'\right] = Ze\,(1 + r/\lambda_D)\exp(-r/\lambda_D) \quad . \tag{2.10}$$

Thus, the charge is $Ze$ at $r = 0$, and decays to 0 for large $r$. At the Debye radius the charge is $Q(\lambda_D) = 2\exp(-1)Ze \sim 0.736Ze$, and falls to half its value at the origin, $0.5Ze$, at $r/\lambda_D \sim 1.68$, well beyond the Debye radius. This result is interesting, when considering $\lambda_D$ as the cut-off impact parameter, when Coulomb collisions are considered.

## 3. Comparison between an accurate (numerical) and the approximate (analytical) solution

For the sake of comparison with Ref. [12], let us normalise the distance $r$ and the potential $\Phi(r)$ in terms of the Debye length, changing to the following dimensionless variables:

$$\rho = r/\lambda_D \quad , \tag{3.1}$$

$$\Psi(\rho) = 4\pi\varepsilon_o\lambda_D\Phi(r)/Ze = (e\Phi(r)/kT)N_D \quad , \tag{3.2}$$

Thus, Poisson's equation (2.1b) for the potential can be rewritten as

$$\frac{d^2}{d\rho^2}[\rho\Psi(\rho)] = -N_D\rho[1 - \exp(\Psi(\rho)/N_D)] \quad , \tag{3.3}$$

whose numerical solution we shall call the "exact" solution, while the normalized approximate solution would then be

$$\Psi_a(\rho) = \frac{\exp(-\rho)}{\rho} \quad . \tag{3.4}$$

valid for

$$\frac{e\Phi(r)}{kT} = \frac{1}{N_D\rho}\exp(-\rho) \ll 1 \quad . \tag{3.5}$$

The main purpose of this paper is to compare the numerical solution $\Psi(\rho)$ to equation (3.3), with the analytic solution (3.4) to the linearized equation.

Indeed, the approximation should be good for large distances, but equation (3.3) also tells us that it will break down for short distances ($\exp(-\rho) \sim 1$) when $N_D \rho \leq 1$. This will happen if $N_D < 1$, in very diluted plasmas, or for short distances, when $r < \lambda_D$. Also, remembering subsection 2.1, note that $N_D \rho = 4\pi n \lambda_D^2 r = (Z/3)(r/r_c)$, which means that for $Z=1$, $N_D \rho > 1$ for $r > 3r_c$. Therefore, the weak plasma approximation is expected to break around

$$\frac{r}{3r_c} \sim \exp(-r/\lambda_D) \sim 1 \qquad (3.6)$$

for short distances.

However, in order to fully understand what all this means, it is important to quantify these statements. The main purpose of this paper is to compare the numerical solution $\Psi(\rho)$ to equation (3.3), with the analytic solution $\Psi_a(\rho)$, (3.4), to the linearized equation.

The numerical solutions are obtained by means of a fourth-order Runge-Kutta routine [13], which is started at the tail of the solution, and integrated backwards, from larger to smaller distances. The initial values $\rho_o$ for the integration can be chosen to be smaller for larger values of $N_D$, because the validity of the approximation breaks at shorter distances. The way in which they were chosen was such that in the first step of integration we satisfy $|\Psi_a - \Psi| < 10^{-11}$. It was found that if this condition was not met, the integration would not be independent of the initial condition, which is something else we sought. This of course, also depends on the step of integration. Another criterion used in order to choose the step of integration was based on the tolerance at which the approximation would break. For this purpose, three different tolerances, $\tau$, were tried: 1, 5, and 10%. By trial and error, the integration steps were chosen in such a way that the distance $\rho_d$ at which $(|\Psi_a - \Psi|/\Psi) \times 100 = \tau$ occurred, would be such that $\tau$ could be determined within a 0.1% uncertainty.

**Table 1**. Initial value $\rho_o$ at which the integration of (3.3) was started, and distance $\rho_d$ at which tolerances of 1, 5, and 10 % between the approximate and exact solutions are reached, for different values of $N_D$.

| $N_D$ | $\rho_o$ | $\rho_d$ 1% | $\rho_d$ 5% | $\rho_d$ 10% |
|---|---|---|---|---|
| 0.1 | 13 | 3.569 | 2.314 | 1.828 |
| 0.3 | 11 | 2.686 | 1.510 | 1.077 |
| 1 | 11 | 1.775 | 0.734 | 0.401 |
| 2 | 11 | 1.288 | 0.368 | 0.137 |
| 5 | 7 | 0.708 | $5.11 \times 10^{-2}$ | $1.94 \times 10^{-2}$ |
| 10 | 5 | 0.340 | $8.43 \times 10^{-3}$ | $7.22 \times 10^{-3}$ |
| 40 | 5 | $1.83 \times 10^{-3}$ | $1.40 \times 10^{-3}$ | $1.34 \times 10^{-3}$ |
| $1 \times 10^2$ | 5 | $5.51 \times 10^{-4}$ | $4.86 \times 10^{-4}$ | $4.71 \times 10^{-4}$ |
| $5 \times 10^2$ | 4 | $8.70 \times 10^{-5}$ | $8.05 \times 10^{-5}$ | $7.87 \times 10^{-5}$ |
| $1 \times 10^3$ | 2 | $4.02 \times 10^{-5}$ | $3.75 \times 10^{-5}$ | $3.68 \times 10^{-5}$ |
| $1 \times 10^4$ | 2 | $3.27 \times 10^{-6}$ | $3.10 \times 10^{-6}$ | $3.06 \times 10^{-6}$ |
| $1 \times 10^5$ | 0.1 | $2.77 \times 10^{-7}$ | $2.66 \times 10^{-7}$ | $2.63 \times 10^{-7}$ |
| $1 \times 10^6$ | 0.01 | $2.41 \times 10^{-8}$ | $2.33 \times 10^{-8}$ | $2.31 \times 10^{-8}$ |
| $1 \times 10^7$ | 0.001 | $2.14 \times 10^{-9}$ | $2.08 \times 10^{-9}$ | $2.07 \times 10^{-9}$ |

As a first result of this calculation we found, is that the approximate solution underestimates the exact one. Table 1 shows, for several values of $N_D$, the initial values $\rho_o$ of the integration, and the values

$\rho_d$ for which tolerances of 1, 5 and 10% between the two solutions fail. The same results are plotted in logarithmic scale in figure 1. For classical, weakly coupled plasmas, relevant $N_D$ values run from 40, for the Solar atmosphere, some gas discharges, and laser produced plasmas, to $10^8$ for the interstellar gas, while for the Solar corona and thermonuclear plasmas, they are around $10^7$ [14]. In addition, we have explored smaller values of $N_D$, including some smaller than one, for which a strongly coupled plasma is expected.

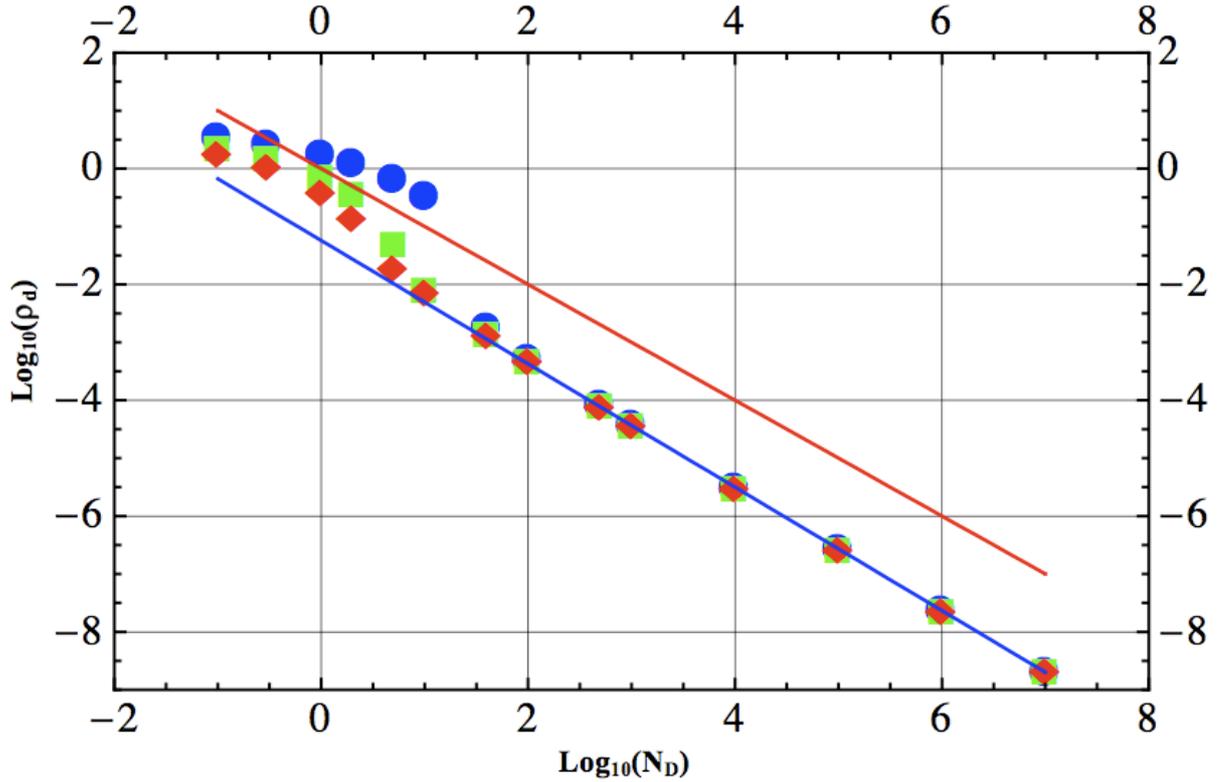

**Figure 1.** (In colour in the on-line version) Distances $\rho_d$ at which the approximate solution breaks, for different values of $N_D$. The blue dots are for the case in which a 1% tolerance is taken, the green squares for 5%, and the red diamonds for 10%. The blue line (below) is the fit for results that go from $N_D = 40$ to $1\times10^7$. The red line (above) is the curve $\log_{10}\rho = -\log_{10}N_D$.

For large values of $N_D$, the distance values at which the approximation breaks down are very similar regardless of the tolerance. In the plot in figure 1, they overlap down to $N_D = 40$. As $N_D$ is further reduced, the distance at which the approximation breaks down increases, and is larger for smaller tolerances. However, it is interesting to observe that the differences obtained for different tolerances tend to diminish for fractional values of $N_D$. While we have followed the calculations into very small values of $N_D$, it should be noted that the description of the charge density, used in the Poisson's equation, in terms of a Boltzmann distribution for such a case may lose its meaning, due to the increasing importance of the temporal fluctuations about the average. Indeed, the use of the Maxwell-Boltzmann distribution assumes the particle species are in equilibrium, which requires the density to be large enough for collisions to allow relaxation to such a state.

Thus, the linear approximation is very good when $N_D$ ranges between 40 and $10^7$, and it is possible to fit the results to the curve $\log_{10}\rho_d = 1.18828 - 1.07493N_D$ (lower line in figure 1), which yields the empirical law

$$N_D^{1.07493}\rho_d = 0.06482 \ . \tag{3.7}$$

This should be compared to the curve $\log_{10}\rho = -\log_{10}N_D$ (upper line in figure 1), which stands for the limit $N_D\rho = 1$. Therefore, our result gives a quantitative meaning to the statement $N_D\rho \ll 1$. We should note that this result differs with that found in Ref. [12]:

$$N_D\rho = 0.01 \tag{3.8}$$

Our work goes further than that of Ref. [12], in that we study the cases of smaller values of $N_D$. Besides, only the case in which the tolerance is 10% was reported. We did not explore $N_D$ as large as in that previous work ($10^{15}$), because they are not relevant for practical purposes, and it is to be expected that the approximation will improve as $N_D$ increases, while we are more interested in exploring when the approximation breaks down.

**4. Conclusions**
In order to study the Debye shielding, the numerical (exact) solution for the electrostatic potential $\Psi(\rho)$ was obtained, from equation (3.3), and it was compared to the approximate solution, given by the Yukawa potential (3.4), found for the linearized approximation. The latter underestimates the exact solution, but is a good approximation for values of $N_D$ as small as 40. Even for $N_D = 5$, the approximation is still good, down to $r = 0.05\lambda_D$ if one requires only a 5% tolerance. The distances $\rho_d$, at which the approximation breaks, given fixed values of $N_D$, for tolerances of 1, 5 and 10% were obtained, and it was observed that they are practically the same, regardless of the tolerance, for $N_D > 40$. Using these data, the empirical law (3.7) was found, which gives a clear quantitative meaning to the statement $N_D\rho \ll 1$, for the validity of the linear approximation.

From the didactical point of view, this result is illustrative the validity of the usual approximation, which is never quantitatively explained in textbooks, even when a Maxwell-Boltzmann equilibrium distribution is assumed, which may not necessarily be right in many cases.

Finally, it must be noted that the results reviewed in this paper are valid for classical plasmas, such that the distance of maximum approach $r_c$, is much greater than the electron's thermal wavelength $\lambda_e = \sqrt{2mkT}$. If $N_D \gg 1$, the electrons become degenerate, and the interference between wave functions must be considered for high temperatures. In this case, the classical model is no longer valid, and the shielding is provided by the Thomas-Fermi length $r_{TF} = (\pi/3n_e)\sqrt{\hbar^2/4me^2}$ (Ref. [8]).

**Appendix  The calculation of the charge density**
The charge density distribution (2.9) can be obtained from Poisson's equation, from the potential (2.8), by appropriately dealing with the singularity. As explained in Classical Electrodynamics textbooks (take for instance Ref. [13]), the potential produced by a point charge $Ze$ at the origin

$$\Phi_p(r) = \frac{Ze}{4\pi\varepsilon_o r} \tag{A.1}$$

satisfies the Poisson equation

$$\nabla^2\Phi_p(r) = -\frac{Ze\delta(r)}{\varepsilon_o} \ , \tag{A.2}$$

where $\delta(r)$ is the Dirac delta function.
In the Debye shielding case, we provide two ways to proceed. The first one, suggested by Greiner in Ref [15], is to sum and subtract the point charge potential in (2.8). Therefore,

$$\nabla^2 \Phi_a(r) = \frac{Ze}{4\pi\varepsilon_o} \nabla^2 \left[ \frac{1}{r} + \frac{\exp(-r/\lambda_D) - 1}{r} \right] = \frac{Ze}{4\pi\varepsilon_o} \left[ -4\pi\delta(r) + \frac{\exp(-r/\lambda_D)}{r\lambda_D^2} \right], \quad (A.3)$$

which yields (2.9), using (2.1a). The second way may be by using the identity of the Laplacian for a product of two scalar functions $\nabla^2(fg) = \nabla^2(f)g + \nabla f \cdot \nabla g + f\nabla^2(g)$, where $f = Ze/(4\pi\varepsilon_o r)$, and $g = \exp(-r/\lambda_D)$. By these means, we get

$$\nabla^2 \Phi_a(r) = \frac{Ze}{4\pi\varepsilon_o} \left[ -4\pi\delta(r)\exp(-r/\lambda_D) + \frac{\exp(-r/\lambda_D)}{r\lambda_D^2} \right], \quad (A.4)$$

which is equivalent to (A.3), since when integration is made to obtain the charge within the sphere of radius $r$, it yields the same result (2.10) for the total charge in a sphere of radius $r$, due to the property of the Dirac delta function

$$\int f(\mathbf{r})\,\delta(\mathbf{r} - \mathbf{r_o})dV = f(\mathbf{r_o}), \quad (A.5)$$

when integrating over a domain containing $\mathbf{r_o}$.

In plasma physics textbooks, this basic electrostatics issue is seldom discussed, although Refs. [3-5,7] do mention the first term in the charge density, but not as a result of the Laplacian of (2.8). We believe it is a good example to stress the importance of properly dealing with singularities.